# Search for Origin of Room Temperature Ferromagnetism Properties in Ni doped ZnO Nanostructure[†]

Amit Kumar Rana[1], Yogendra Kumar[1], Parasmani Rajput[2], S. N. Jha[2], D. Bhattacharyya[2], Parasharam M. Shirage[1,*]

[1]Discipline of Physics & Metallurgical Engineering and Materials Science, Indian Institute of Technology Indore, Khandwa Road, Simrol Campus, Indore 453552, India.

[2]Atomic & Molecular Physics Division, Bhabha Atomic Research Centre, Trombay, Mumbai 400085, India.

[*]Author for correspondence E-mail: paras.shirage@gmail.com , pmshirage@iiti.ac.in

[†]Supporting information

## Abstract

The origin of room temperature (RT) ferromagnetism (FM) in $Zn_{1-x}Ni_xO$ (0<x<0.125) samples are systematically investigated through physical, optical, and magnetic properties of nanostructure, prepared by simple low-temperature wet chemical method. Reitveld refinement of X-ray diffraction pattern displays an increase in lattice parameters with strain relaxation and contraction in Zn/O occupancy ratio by means of Ni-doping. Similarly scanning electron microscope demonstrates modification in the morphology from nanorods to nanoflakes with Ni doping, suggests incorporation of Ni ions in ZnO. More interestingly, XANES (X-ray absorption near edge spectroscopy) measurements confirm that Ni is being incorporated in ZnO as $Ni^{2+}$. EXAFS (Extended X-ray Absorption Fine Structure) analysis reveals that structural disorders near the Zn sites in the ZnO samples upsurges with increasing Ni concentration. Raman spectroscopy exhibits additional defect driven vibrational mode (at 275 $cm^{-1}$), appeared with Ni-doped sample only and the shift with broadening in 580 $cm^{-1}$ peak, which manifests the presence of the oxygen vacancy ($V_O$) related defects. Moreover, in photoluminescence (PL) spectra we observed peak appears at 524 nm, indicates the presence of singly ionized $V_O^+$, which may activate bound magnetic polarons (BMPs) in dilute magnetic semiconductors (DMSs). Magnetization measurements indicate weak ferromagnetism at RT, which rises with increasing Ni consolidation. It is therefore proposed that effect of the Ni-ions as well as the inherent exchange interactions rising from $V_O^+$ assist to produce BMPs, which are accountable for the RT-FM in $Zn_{1-x}Ni_xO$ (0<x<0.125) system.

# 1. Introduction

One of the new generation semiconductors known as DMSs, in which small fraction of the non-magnetic site *i.e.* the host cations are replaced by magnetic ions. DMSs have drawn much attention to scientific community because of their promising technological applications in the emerging field of spin electronics devices *i.e.* in spintronic devices, such as non-volatile memory[1], spin light emitting diode[2], spin based quantum computers, spin field effect transistors, logic devices[3] *etc*. These spintronic devices are multifunctional devices with superior efficiency, lower power consumption and higher speed. The elementary requirement for application point of view, the Curie temperature ($T_C$) of DMS material is 300 K or beyond. Recently researchers are focusing on metal-oxide-based DMSs, such as ZnO, $TiO_2$, $CeO_2$, $SnO_2$ *etc*., among these possible, ZnO has been widely studied as a host materials for DMS. It is because ZnO was predicted theoretically to be a capable to attain RT DMS. In addition to this, it is also recognized as semiconductor with large excitonic binding energy (60meV) [4], wide-band gap (3.37eV) and exhibits multifunctional properties. The idea of DMS with dilute doping of magnetic elements in a host semiconductor to create the RT magnetic semiconductor was first predicted by Dielt *et al.*[5] on the *p*-type wide band gap semiconductor GaN and Mn doped ZnO. Similarly Sato *et al.*[6,7] by theoretically calculation, based on local density approximation shows ferromagnetic ordering with $T_C$ above room temperature for *n*-type ZnO with doping of V, Cr, Co, Fe and Ni. Above predication of Dielt *et al.* and Sato *et al.* and results of last few decades have displayed that doping of transition metal (*TM*) elements such as Cu, Ni, Co, Fe, Mn, *etc.* in a wide-gap semiconductors provides a conceivable means of tuning of both optical properties as well as ferromagnetism properties of a single material[8-11]. It also demonstrates that by simple change in the shape and size of nanomaterial it is possible to tune the physical properties of nanostructure [12-14]. However, present and existing theories and results[5-7] cannot satisfactorily articulate the origin of the witnessed ferromagnetism. Recently, RT-FM was detected in undoped in addition to nonmagnetic element doped in semiconductors. For example Schwartz *et al.*[15] have detected strong ferromagnetic behavior with $T_C$ above 350K in Ni doped ZnO films. Meanwhile Wakano *et al.*[16] have noticed ferromagnetism at very low temperature of 2K and super-paramagnetic at 30K for Ni (with 25%) doped ZnO while Cong *et al.*[17] detected ferromagnetism with $T_C$ above 335K in $Zn_{1-x}Ni_xO$ (x=0.03) nanoparticles. However, Ueda *et al.*[18] have reported that Co- doped in ZnO films exhibit FM above RT, though Ni-, Cr- and Mn- doped ZnO films do not display any indication of ferromagnetic behavior. Yin *et al.* have also verified that there is no ferromagnetism observed in Ni doped ZnO films[19]. B. B.

Straumal *et al.*[20-21] substantially provided the evidence of grain boundary effect on the FM performance of virgin and *TM*-doped ZnO. These results unveils about the mechanisms of RT- FM in Ni doped ZnO is still an open controversial issue and whether FM in *TM*- doped ZnO is an inherent or extrinsic property is a topic of debate. To ascertain the origin of ferromagnetism ordering in Ni doped ZnO DMS nanostructures, we have attempted to present a systematic study of Ni doping on local crystal structural, optical and magnetic properties of ZnO nano-crystalline samples. We have selected wet chemical method for the synthesis of nanostructures as it involves low temperature processing, cost effective and deals with a higher degree of solubility. Crystal structural and local structure around Zn ions in Ni-doped ZnO nanostructures have been investigated by X-ray diffraction (XRD), EXAFS and Raman spectroscopy. Optical and magnetic properties of the samples have been studied by photoluminescence spectroscopy and vibrating sample magnetometer (VSM) measurements, respectively. It has been found that the incorporation of Ni in ZnO nanostructure not only changes its lattice constant, but also produces oxygen and Zn related defects, which alters nanostructure morphology, optical properties and also improves ferromagnetic properties of the sample, which is crucial to develop an optical spintronic device and high-density magnetic storage media.

## 2. Experimental Details

$Zn_{1-x}Ni_xO$ ($0<x<0.125$) samples were synthesized by simple wet chemical process have been named as ZnO, ZnO: Ni5%, ZnO: Ni7.5%, ZnO: Ni10%, ZnO: Ni12.5% for different Ni-concentration x = 0, 5%, 7.5%, 10%, and 12.5%, respectively. We have used appropriate amounts of analytical grade metal nitrates like Zinc Nitrates hexa-hydrate, Nickel Nitrate hexa-hydrate, (100 mM: Alfa Aesar chemicals) powders which were thoroughly mixed and dissolved in double distilled water and stirred for 30 min, subsequently ammonia was added and pH of the solutions were maintained at ~11. Then the solutions were heated at 85 °C for 2 hrs. and solutions were filtered using Whatman's filter paper and washed numerous times with double distilled water. Subsequent to overnight drying pure and Ni doped ZnO samples were collected and annealed at 150°C for 2 hrs. to obtain phase pure samples. The crystal phase and surface morphology of $Zn_{1-x}Ni_xO$ samples were investigated by X-ray diffraction (XRD), using Advanced X-ray diffractometer Bruker D8 with Cu-Kα radiation and Supra-55 Zeiss is use for FESEM (field emission scanning electron microscope). X-ray absorption spectroscopy (XAS) measurements, which include both Extended X-ray Absorption Fine Structure (EXAFS) and X-ray absorption near edge spectroscopy (XANES) measurements of Zn and Ni K-edges, were carried out at the beamline-9 at the Indus-2 synchrotron source at the RRCAT (Raja Ramanna Centre for Advanced Technology) Indore, India. XANES and EXAFS measurements were done in transmission mode using ionization chambers. The beamline consists of Rh/Pt coated meridional cylindrical mirror for collimation and Si (111) based double crystal monochromator to select excitation energy. The energy range of XAFS was calibrated using Zn and Ni foils at 9659 and 8333 eV. The EXAFS data has been examined using FEFF 6.0 code[22], which comprises Fourier transform with background reduction to derive the versus R spectra from the absorption spectra using ATHENA software, generation of the theoretical EXAFS spectra starting from a presumed crystallographic structure and lastly fitting of experimental data with the theoretical spectra using ARTEMIS software[23]. Room temperature fluorescence spectroscopic measurement was conducted using a spectrofluorometer having *Xe* lamp source (excitation wavelength of 325 nm), Horiba Jobin Yuon fluorolog-3. Raman scattering measurement of the samples was carried out using Labram-HR 800 spectrometer with excitation radiation (wavelength of 488 nm) from an argon ion laser at a spectra with resolution of 1cm$^{-1}$. Lakeshore (model no. 7407) VSM (Vibrating sample magnetometer) was used for D.C. magnetization measurements of the samples.

## 3. Results and discussion

The Reitveld refined XRD patterns of $Zn_{1-x}Ni_xO$ (0<x<0.125) are represented in figure 1(*a*). It displays that all peaks of $Zn_{1-x}Ni_xO$ (0<x<0.125) samples resemble with the standard Bragg positions for hexagonal phase of wurtzite ZnO (space group *P6₃mc*), which have been indicated by the upright lines at the bottom of the XRD patterns. Ni doping does not lead to the occurrence of additional peak or vanishing of peak related to hexagonal wurtzite pure ZnO structure in XRD pattern and confirms that the samples are phase pure without any change in their wurtzite phase. There is no trace of impurity related to NiO or and any binary phases of zinc/nickel are observed up to 12.5% Ni doping. However the probability of existence of the minute phase or small cluster cannot be completely ruled out considering the detection limit of XRD technique. It has been noticed from *fig. 1(a),* that all the sample show strong directional growth along (002) plane, which indicates that growth along *c*-axis is more preferred as compared to other axis**.** This might be due to the nucleation of rods like morphology of the samples, which is most common in ZnO.

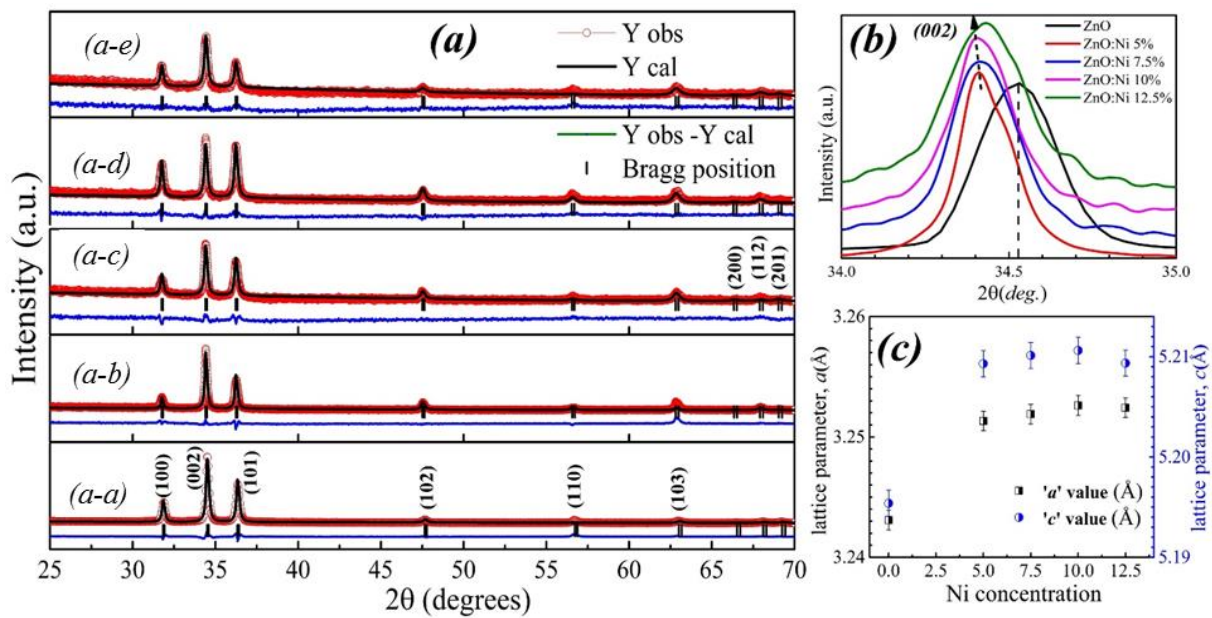

***Figure 1.*** *(Colour online) Shows the **(a)** Reitveld refined XRD patterns of (a-a) ZnO (a-b) ZnO: Ni 5% (a-c) ZnO: Ni 7.5% (a-d) ZnO: Ni 10% and (a-e) ZnO: Ni 12.5%. **(b)** Displays the shift in 2θ after Ni doping in the diffraction peak corresponds to (002) peak **(c)** Displays the variation in lattice parameter "a" and "c" (in unit Å) of pure and Ni doped ZnO.*

However with an increase in Ni concentration, there is a signature of growth along (100) and (101) directions also, which gives the hint of alteration in morphology or orientation and incorporation of Ni into ZnO lattice[24]. Due to comparable ionic size of $Ni^{2+}$ (0.55Å) and $Zn^{2+}$ (0.60Å), Ni ions replace $Zn^{2+}$ ions easily in ZnO lattice, without distorting the crystal structure significantly. However, vigilant investigation of the peak location of samples show that the *(002 )* peak is relocate towards lower two theta value up to 10% Ni doping and in higher doping concentrations, the peak is slightly rearrange to higher two theta values, as shown in *fig. 1(b)*. Form the Reitveld refinement, full detail of the lattice parameter and crystal structure of $Zn_{1-x}Ni_xO$ (0<x<0.125) samples are given in *Table S1*. It shows, Zn/O occupancy ratio is less than 1, moreover with Ni doping it decreases, therefor this support the defect formations mechanism is due to doping[24]. The lattice parameters computed from the Reitveld refinements are presented in *fig.1(c)* with respect to Ni doping. Fig.1(c) provides evidence of an abrupt escalation in lattice parameter because of Ni doping followed by a gradual increase up to 10% doping. A very slight decline in *'a'* and *'c'* lattice constants can be evidenced in higher Ni doping (12.5%). The above results, which are similar to previous observations[25, 26] where defects generated by $Ni^{2+}$ ions and manifests successful incorporation of $Ni^{+2}$ at the $Zn^{2+}$ sites of the ZnO lattice. Unit cell Volumes *(V) for* a hexagonal structure and average bond lengths ($L$) of Zn-O have been calculated from using eqn. (1) and (2) are shown in fig. 2($a$).

$$V = 0.866 \times a^2 \times c$$ ------------------- (1)

and

$$L = \sqrt{\frac{a^2}{3} + (0.5 - u_p)^2 c^2}$$ ------------------- (2)

Where "$u_p$" is the positional parameter, defined as the bond length parallel to the $c$-axis, along the '$c$' axis[27] (given by $u_p = \frac{a^2}{3c^2} + 0.25$). Fig. 2(b) displays the uniform tensile micro strain which is also one of the reasons for the shift of the Bragg's peak towards lower $2\theta$ value up to 10% Ni doping concentration and it is observed that higher Ni doping creates uniform compressive strain with an increase in $2\theta$ value. The lattice strain in the crystal structure is calculated by Williamson–Hall methods have been already discussed in detail elsewhere[24]. It is clearly visible that the increase in lattice parameter has strong correlation with decrease in strain in the nanostructures or vice versa. Similar results have been described by Kumar $et$ $al$.[28] in case of Co-doped ZnO and Dakhel $et$ $al$.[29] for Gd-doped ZnO where such change in lattice parameter with change in strain value has been recognised with the creation of oxygen related defect such as oxygen vacancies. Thus from XRD results we conclude that Ni[2+] ions get successfully incorporated in the host structure and produce oxygen vacancies in the sample. Besides this variation in lattice parameters also illustrate the distortion in crystal structure because of Ni[2+] ions doping into the ZnO crystal lattice. The degree of distortion shown in fig. 2($c$) is calculated by the following relation:

$$R = \frac{2a(2/3)^{1/2}}{c}$$ ------------------------------------ (3)

Where $R = 1$ gives the ideal wurtzite structure[30], here '$a$' and '$c$' are lattice constants computed from the Reitveld refinement[31].

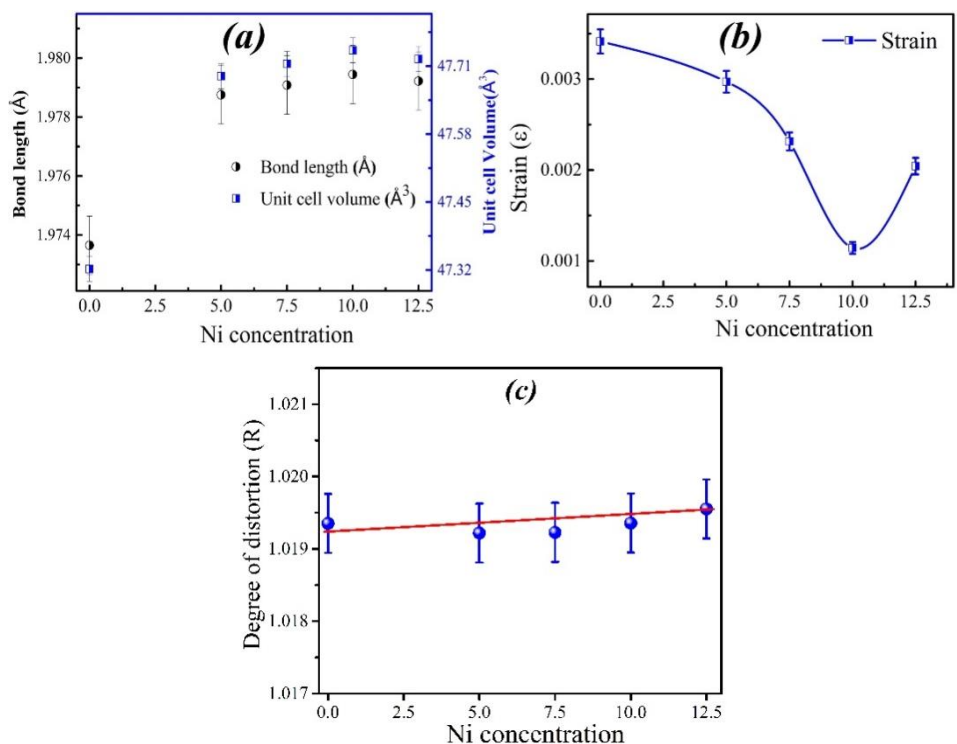

**Figure 2.** (Colour online) (**a**) Shows the change in the average bond length (in Å) and unit cell volume (in Å³), (**b**) displays the crossponding change in micro Strain (ε) value, and (**c**) represents the change in the R (degree of distortion) with increasing the Ni concentration.

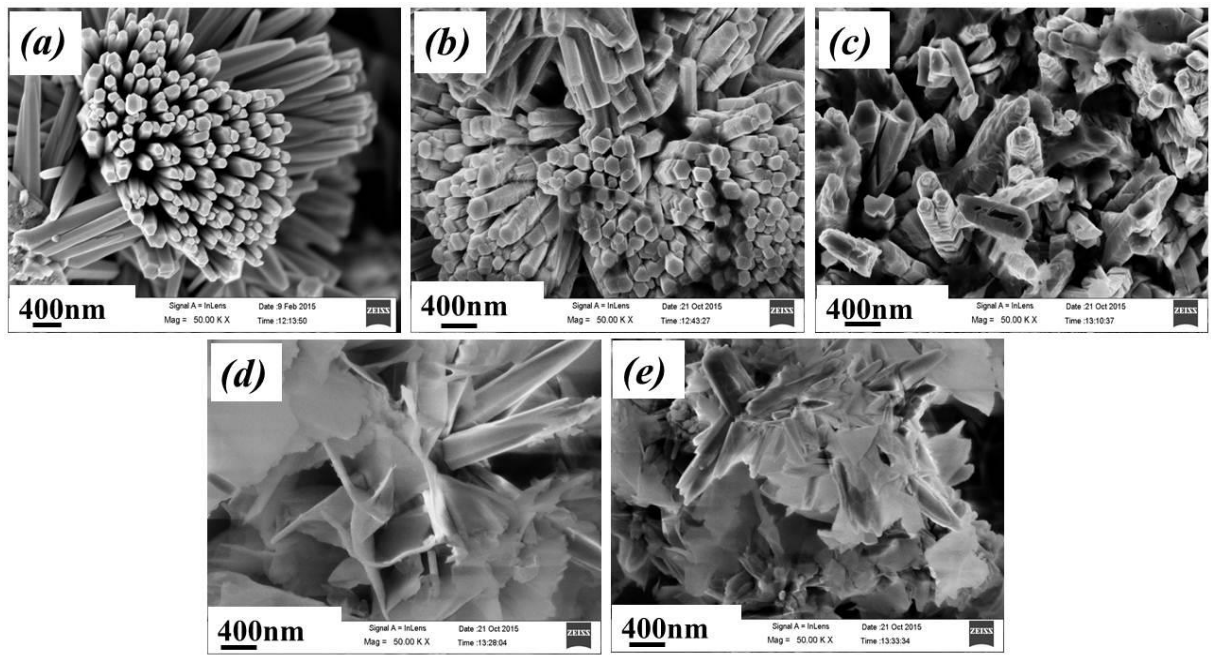

**Figure 3.** Represents the FESEM images of (**a**) ZnO: Ni 0%, (**b**) ZnO: Ni 5%, (**c**) ZnO: Ni 7.5%, (**d**) ZnO: Ni 10%, and (**e**) ZnO: Ni 12.5% (magnification at 50 KX).

The surface morphology of the nanostructure has been observed by FESEM. Fig.3 displays the SEM image (magnification at 50 KX) of pure and Ni doped ZnO nanostructure. In case of ZnO nano-rods are observed and with highest 12.5 % Ni doped ZnO samples displays nano-flakes like nanostructure morphology along with nanorods. It is clearly visible from *fig.3*, that Ni doping increases the size of ZnO nanostructures, in case of pure ZnO there is rod like morphology with diameter ~100 nm and length ~2 $\mu$m and with 5%, 7.5% and 10% Ni doping the diameters changes to ~150 nm, ~200 nm and ~300 nm, respectively. However in case of 10% and higher doping concentration of Ni, flake-like morphology also co-exists along with rods. This suggests that Ni is replacing Zn in ZnO and altering its morphology. From the EDXS (energy dispersive X-ray spectra) of all samples, the atomic percentages of Ni and ZnO in the samples are estimated and shown in Table S2.

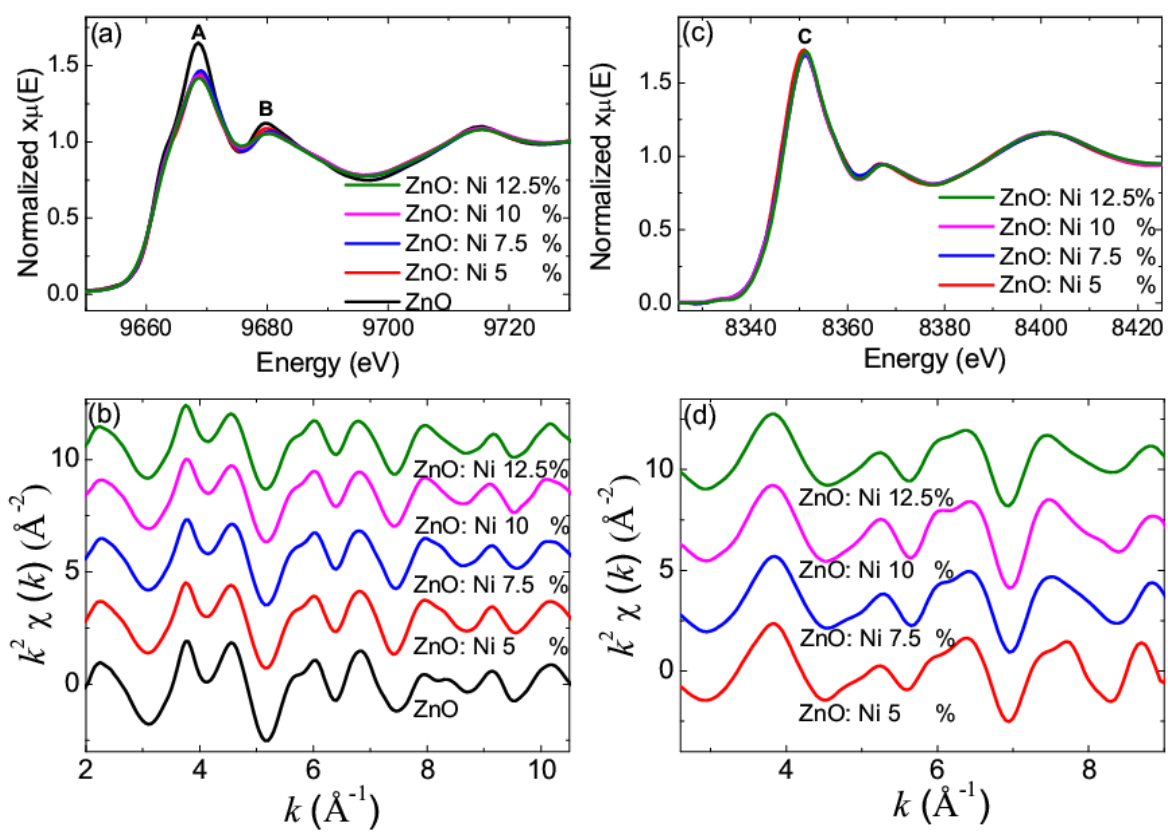

**Figure 4.** *(Colour online) Ni doped ZnO nanostructures (**a**) Normalized XANES data at Zn K-edge, (**b**) $k^2$- weighted $\chi(k)$ spectra at Zn K-edge, (**c**) Normalized XANES data at Ni K-edge, (**d**) $k^2$- weighted $\chi$ (k) spectra at Ni K-edge. The $\chi$ (k) data has been vertically shifted for clarity.*

Figure 4 (*a, c*) represents normalized XANES spectra and Figure 4 (*b, d*) shows $k^2$- weighted $\chi(k)$ XAFS spectra of Ni doped ZnO nanostructure at Zn and Ni *K*-edges. One may note that there is variation of white line intensity (denoted as peak A) and intensity of peak B for Ni doped ZnO nanostructure. There is huge drop in the intensity of peak A & B of Zn K-edge XANES whereas a very small change at peak C of Ni edge XANES, which mainly arises due to variation of charge-transfer[32-34] upon Ni[2+] doping in ZnO nanocrystal. The variation in the white line intensity suggests that there is large variation in number of vacancies/holes due to Ni doping. For detailed analysis of local structure Fig. 5(*a, b*) shows Fourier transform (FT) of EXAFS spectra of samples. The *k*-range of 2.5-10 Å[-1] and 2.5-9 Å[-1] were used for FT of Zn and Ni EXAFS data, respectively. For Zn *K*-edge EXAFS fitting the structure is assumed as pure ZnO wurtzite, where Zn is co-ordinated with four O atoms (Zn-O) at 1.98Å distance in first shell and second next near-neighbour Zn is bounded by 12 Zn atoms (Zn-Zn) at 3.21 Å distance. The fittings were performed in phase un-corrected R-space range of 1-3.21 Å. Similarly, the Ni *K*-edge data were fitted assuming that Zn atoms are replaced by Ni atoms in wurtzite structure. During fitting of the data, coordination numbers of the different shells, Debye-Waller factor ($\sigma^2$) and bond distance (*R*) were fitted as free parameters whereas non-structural parameter $E_0$ was fixed Figure 5(*a, b*) also shows the best fitted curves and best fitting values of the parameters are listed in Table 1 and Table 2.

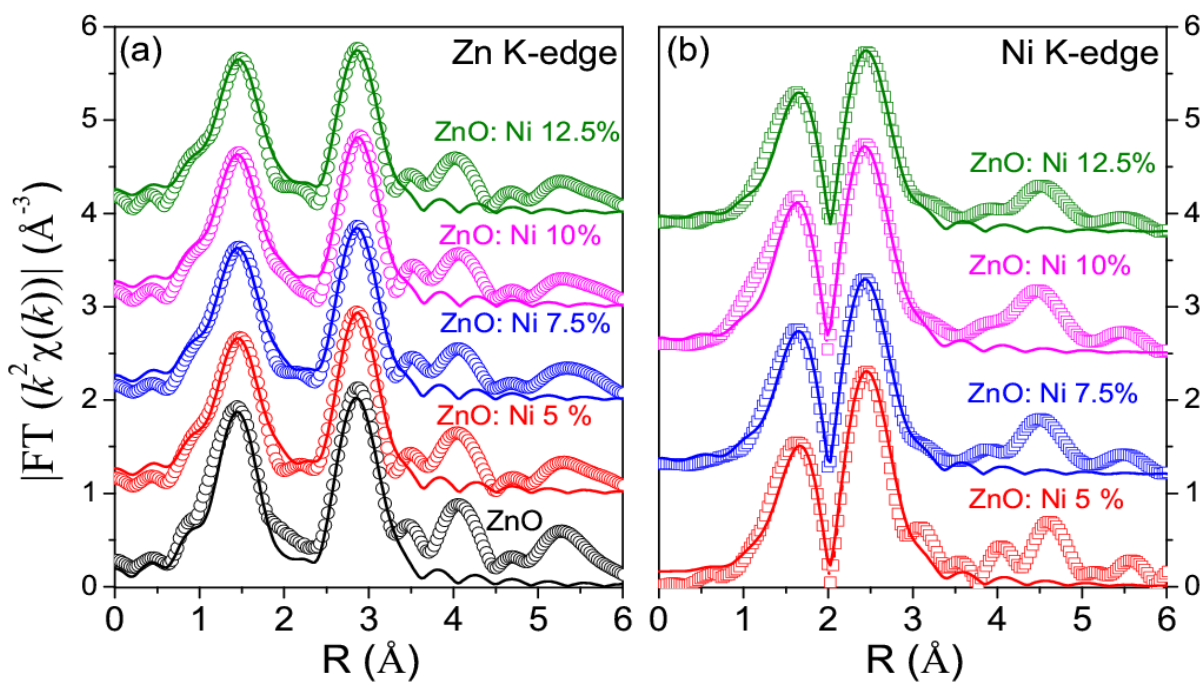

**Figure 5.** *(Colour online) Fourier transform of $k^2$-weighted of (**a**) Zn K-edge, (**b**) Ni K-edge, for Ni doped ZnO nanostructure. The symbol shows empirical data and solid lines are the best fitting data. The curves are vertically shifted for clarity.*

Results of Zn K-edge EXAFS fit reveal that Zn-O bond distance is slightly reduced on Ni doping while no change in Zn-Zn bond distance is observed. Ni K-edge fitting however shows decrease in both Ni-O and Ni-Ni/Zn bond distances with rise in Ni substitution concentration. Although the ionic radii of $Zn^{+2}$ and $Ni^{+2}$ are almost same (0.60 and 0.55 Å respectively), the change in bond lengths are insignificant in this case. Nonetheless, there is a systematic increase in Debye-Waller factor ($\sigma^2$) (DWF) values for both Zn-O and Zn-Zn paths which indicate an increase in local disorder in ZnO lattice near the host sites due to Ni doping. DWF values of Ni-O and Ni-Ni/Zn paths however do not show any change. Similar increase in Debye-Waller factors at site of Zn atom have also been detected by us in case of Co and Fe doped ZnO nano-crystalline samples also[32, 33]. It is noticeable that there is systematic decrease in coordination numbers (CNs) of Zn-O, Zn-Zn and Ni/Zn-Zn whereas no variation in CN of Ni-O. This confirms that with Ni doping number of holes/vacancies were increased in Ni doped ZnO nanocrystals as it has been observed from XANES also.

**Table 1**. *From the Zn K-edge EXAFS data fitting, we have obtained variation of CN (co-ordination number), R (bond distance) and $\sigma^2$ (Debye-Waller factor). The amplitude reduction factor ($S_0^2$) and $\Delta E$ were obtained from pure ZnO fitting 1.05±0.08 and 1.9±0.5, respectively and kept fixed for all the samples. The numbers in parentheses indicate the uncertainty in the last digit.*

| Sample | $CN_{Zn-O}$ | $R_{Zn-O}$(Å) | $\sigma^2_{Zn-O}$ (Å$^2$) | $CN_{Zn-Zn}$ | $R_{Zn-Zn}$(Å) | $\sigma^2_{Zn-Zn}$ (Å$^2$) |
|---|---|---|---|---|---|---|
| ZnO | 3.7(3) | 1.952(2) | 0.0040(3) | 11.8(3) | 3.211 (3) | 0.0088 (2) |
| ZnO:Ni5% | 3.3(2) | 1.952(3) | 0.0055(2) | 11.5(2) | 3.211 (5) | 0.0091(3) |
| ZnO:Ni7.5% | 3.3(2) | 1.949(2) | 0.0056(2) | 11.4(2) | 3.208 (6) | 0.0094(2) |
| ZnO:Ni10% | 3.0(3) | 1.943(3) | 0.0056(2) | 11.3(3) | 3.209 (5) | 0.0095(2) |
| ZnO:Ni12.5% | 2.8(2) | 1.942(2) | 0.0062(3) | 10.9(2) | 3.209 (6) | 0.0108(2) |

**Table 2**. *Structural parameters CN (co-ordination number), R (bond distance) and $\sigma^2$ (Debye-Waller factor) obtained from Ni K-edge EXAFS data fitting. Similar to Zn K-edge, the amplitude reduction factor ($S_0^2$) and $\Delta E$ were obtained from pure NiO fitting 1.01±0.06*

*and 1.8±0.6, respectively and kept fixed for all the samples. The numbers in parentheses*
*indicate the uncertainty in the last digit.*

| Sample | $CN_{Ni-O}$ | $R_{Ni-O}$ (Å) | $\sigma^2_{Ni-O}$ (Å$^2$) | $CN_{Ni-Ni/Zn}$ | $R_{Ni-Ni/Zn}$ (Å) | $\sigma^2_{Ni-Ni/Zn}$ (Å$^2$) |
|--------|-------------|----------------|---------------------------|------------------|--------------------|-------------------------------|
| Ni 5% | 3.9(3) | 2.111(4) | 0.0030(2) | 10.7(3) | 2.944(3) | 0.0082(3) |
| Ni 7.5% | 3.8(2) | 2.106(6) | 0.0030(3) | 10.5(2) | 2.932 (4) | 0.0087(3) |
| Ni 10% | 3.9(2) | 2.092(4) | 0.0030(2) | 10.6(3) | 2.936(3) | 0.0084(3) |
| Ni 12.5% | 3.9(2) | 2.094(3) | 0.0030(2) | 9.9(2) | 2.937(6) | 0.0094(2) |

Raman spectroscopy is very sensitive and one of the powerful techniques for study the change in local structure, defect state and disorder in the ZnO host lattice because of incorporation of transition metal ions[35]. Besides, this technique is used for study of crystalline quality of nanostructure. Wurtzite ZnO has 4 atoms in their unit cell and belonging to $P6_3mc$ symmetry space group, in which total 12 number of phonon mode are present, six TO (transverse optical) mode, two TA (transverse acoustic) mode, three LO (longitudinal optical) mode, and one LA (longitudinal acoustic) mode. The Γ-point of the Brillouin zone, the complex illustration of optical phonons are represented by $\Gamma_{opt}$= A$_1$+2B$_1$+E$_1$+2E$_2$, where E$_1$ and A$_1$ mode are polar in nature and can divided into longitudinal and transverse phonons and all are infrared and Raman actives. E$_2$ modes are non-polar and Raman active, however B$_1$ modes are silent or Raman inactive. The lattice vibrations with A$_1$ and E$_1$ modes, the atoms move parallel and perpendicular to the *c*-axis respectively.

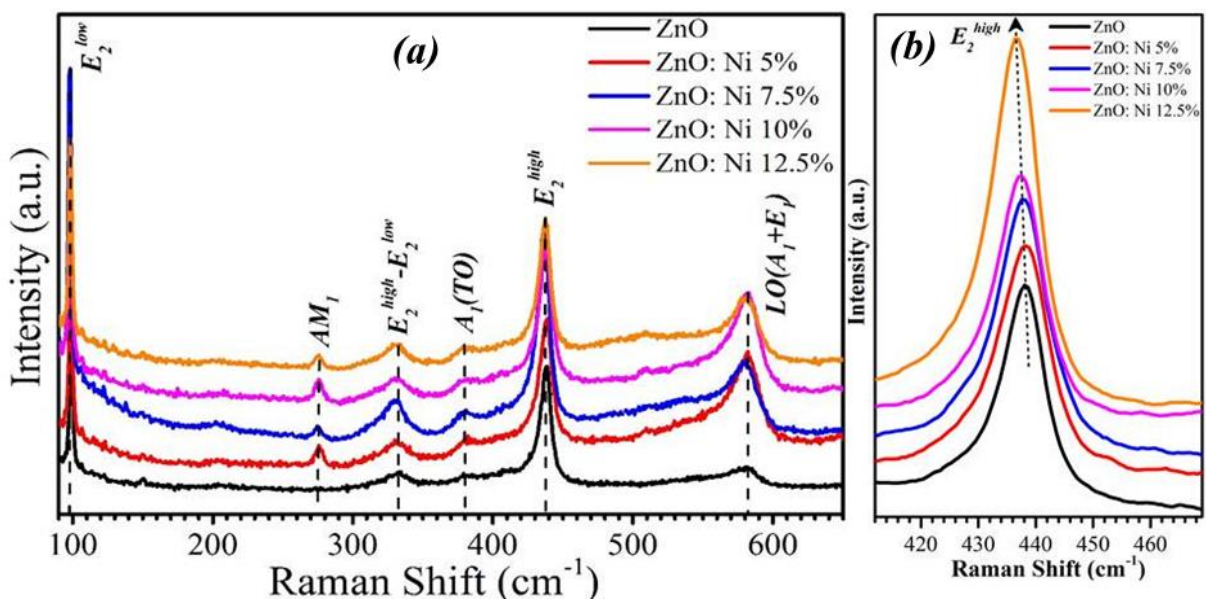

**Figure 6.** *(Colour online) Displays (a) RT Raman spectra of pure and Ni doped ZnO (b) The Raman shift in the $E_2^{high}$ peak.*

Sharpest and strongest peaks at $99 cm^{-1}$ and $438 cm^{-1}$ presented in *fig.* 6 are allocated to the high and low frequency branch of $E_2$ mode, $E_2^{high}$ and $E_2^{low}$ respectively. $E_2^{low}$ is due to the vibration of weighty atom *i.e.* Zn and $E_2^{high}$ is due to Oxygen sub-lattice, which are the characteristic and most prominent peak of wurtzite ZnO structure[36, 37]. The modes, $A_1(TO)$ and $E_1(TO)$ show the polar lattice bond strength[38] while $A_1$ (LO) and $E_1$ (LO) modes can be detected in the unpolarised Raman spectra under back scattering geometry in a bulk ZnO. However, when the size of crystal is in nanometre level, first-order Raman scattering is relaxed (for selection rule with $k = 0$) and phonon scattering is not restricted to the centre of the Brillouin zone[39]. As a result of this, the ZnO nanostructure shows different Raman modes ($E_2^{high}$, $E_2^{low}$, $A_1(LO)$, $A_1(TO)$, $E_1(LO)$, and $E_1(TO)$ modes). RT Raman spectra, range from wavenumber 90 to $700 cm^{-1}$ is presented in *fig.*6 which shows all prominent peak of wurtzite ZnO after Ni doping with additional peak at $276 cm^{-1}$ appearing in Ni doped sample only (*See supporting figure S1*). As the Ni content increases, some of the Raman modes appear broad and highly intense without appreciable shift in comparison to pure ZnO. This provides the hint of change in local symmetry of the nanostructures due to incorporation of Ni atom in the host lattice, while the crystal structure remains the same. This fact is corroborated from EXAFS result also as discussed earlier. Table-3 summarizes the details of the peaks appearing in $Zn_{1-x}Ni_xO$ (0<x< 12.5) samples. We have already discussed above the origin of $E_2^{high}$ and $E_2^{low}$, which are the representative peaks of wurtzite structure. It is observed from Table-3 that with increase Ni-concentration, there is a very small redshift in non-polar mode $E_2^{high}$ in contrast to pure ZnO. This result can be ascribed to the fact that $Ni^{2+}$ substitution prompts the atomic structural irregularities in the periodic Zn atomic sublattice and interrupts its translational symmetry with increase in local alterations in the lattice. This disorder and local distortion interrupt the long-range order in ZnO and deteriorates the electric field related with a mode[40]. The peaks at 333 cm$^{-1}$ and 383 cm$^{-1}$ corresponds to the $E_2^{(high)}$–$E_2^{(low)}$ and $A_1$(TO) respectively, which is due to multi phonon process and can be ascribed the single crystalline nature of ZnO[39]. In comparison with earlier reports of ZnO, two LO modes are located at 587 cm$^{-1}$ ($E_1$) and 575 cm$^{-1}$ ($A_1$) [41, 42]. These LO modes arise due to defect such as $V_O$, $Zn_{in}$ or free carriers, present in ZnO nanostructure[43]. Transition metal (such as Fe and Mn) replacement in ZnO, shows significant modification in $E_1$(LO) mode because of variation in inherent host lattice defects[40, 41]. In our case, Raman scattering peak is centred around 580cm$^{-1}$ for pure ZnO and Ni doping this peak becomes broad and slightly shifted to 583cm$^{-1}$. As a result, this peak can be allocated to the mixture of both LO ($A_1$ and $E_1$) vibrational modes of ZnO. Apart from these five foremost peaks, we have observed an additional mode (AM) only in doped samples at ~276 cm$^{-1}$ (*see supporting Figure S1*). So far the source of this peak is still unclear, as the previous literature ascertains dissimilar reasons for the existence of this peak. Kaschner *et al.*[44] reported this AM in Raman spectra at 274 cm$^{-1}$ associated with Nitrogen (*N*) - doping; appealing that with *N* substitution the intensity of this peak growth and have assigned them as nitrogen-induced local vibration modes (LVMs). Similarly Wang *et al.*[45] have also reported this in *N*- doped ZnO sample. However, later similar group investigated that these additional modes also appears in the Raman spectra of *Fe-, Sb-,* and *Al*-doped ZnO films, and proposed for the inherent host lattice defects[41]. Also, Du *et al.* in (Co, Mn) co-doped ZnO film; they correlated this peak with the oxygen vacancy and Zn interstitial[46]. Ye *et al.*[47] have proposed two plausible mechanisms to describe the source of this AM: LVMs (Local Vibrational Modes) and DARS (Disorder Activated Raman Scattering). The DARS is prompted by the collapse of the translation symmetry of the lattice caused by either impurities or defects due to variation in the growth condition or dopant nature. Therefore, it can be expected, AM in Ni doped samples could rise because of both or either of aforesaid two mechanisms. Thus from Raman study, we can conclude that Ni doping in ZnO nanostructures leads to generate a disorder and lattice defects by troubling the long range ionic ordering in the host ZnO nanostructure.

**Table 3.** *Phonon mode frequencies of wurtzite $Zn_{1-x}Ni_xO$ (0<x<0.125) nanostructure (in units of $cm^{-1}$).*

| Vibration frequency  (cm⁻¹) | | | | | Assignment |
|---|---|---|---|---|---|
| ZnO | ZnO:Ni 5% | ZnO:Ni7.5% | ZnO:Ni 10% | ZnO:Ni 12.5% | |
| 99 | 98 | 98 | 98 | 99 | $E_2^{low}$ |
| - | 276 | 274 | 276 | 276 | **AM** |
| 333 | 331 | 331 | 330 | 333 | $(E_2^{high} - E_2^{low})$ |
| 383 | 382 | 380 | 382 | 382 | $A_1$ **(TO)** |
| 439 | 439 | 437 | 437 | 436 | $E_2^{high}$ |
| 580 | 581 | 581 | 582 | 583 | **LO ($A_1 + E_1$)** |

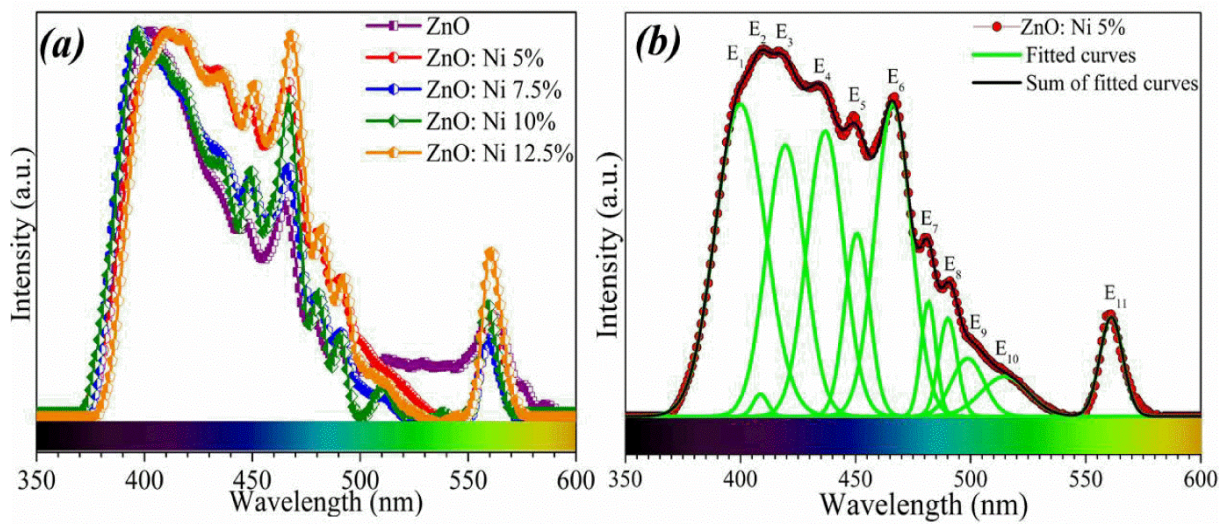

**Figure 7.** *(Colour online)* **(a)** *Demonstrates the RT Photoluminescence spectra of $Zn_{1-x}Ni_xO$ (0<x<12.5) and* **(b)** *Displays RT- PL spectra of Ni 5% doped ZnO. The red dots display experimental data, green lines Gaussian fitting and black line sum of all fitted peaks.*

Photoluminescence (PL) spectroscopy is one of the most sensitive non-destructive and very effective optical approaches to describe the existence of the intrinsic and extrinsic defects in a sample. It's not only provides the information regarding energy state of defects and impurities at a very low densities but also provides useful information regarding structural defects present in semiconductors which plays a crucial role in the growth of RT FM ordering in ZnO based DMSs. Figure 7(a) displays the RT PL spectra of $Zn_{1-x}Ni_xO$ (0<x<0.125) samples measured by exciting wavelength of 325 nm with the energy range from 350 nm to 600 nm. Detected broad PL spectra in the UV-visible region indicate the existence of multi-component contributions which is extracted with the fitting of multiple Gaussian peaks and shown in fig. 7(*b*) (See supporting information Figure S2).

The characteristic wide-range emission spectrum of pure ZnO ranging from near band edge (NBE) (380nm) to yellow emission (600 nm) and this can be well committed into ten peaks at 396, 412, 431, 437, 450, 467, 481, 492, 524, and 563 nm. Ni doping in ZnO produces a shift and additional defect as compare to pure ZnO (see supporting information Table S3). The first peak ($E_1$) at 396 nm is related to NBE, which arises due to the recombination of free exciton existing in the ZnO nanostructure[48]. Ni doping shows the red shift in NBE for Ni 5% doping; this red shift can be feature of robust exchange interaction between *s-p* electrons and d-electrons of $Ni^{2+}$ ions of ZnO band[49]. $E_2$ peak at 401 nm is attributing to the first longitudinal optical phonon copy of free excitons[50] while an origin of rest of the peaks has been discussed in detail in our previous study[24, 51]. We also observed there is change in the relative intensity after Ni doping; it is due the dependence of intensity on the concentration of electrons at particular defect level.  After Ni doping the concertation of electron increases and creates a defect level near the conduction level. This might be one of the reasons for change in the relative intensity after doping (see supporting information Table S3). There is a broad and intense green emission peak $E_{10}$ is also observed in all ZnO samples (see supporting figure S2). The cause of this peak is considered as a green emission (504-524nm) is due to the recombination of photo generated hole and the electron trapped by single ionized oxygen vacancies ($V_o{}^+$), this proof of presence of singly ionized oxygen vacancy states[51, 52]. However oxygen vacancies, which are an important class of point defects in oxides and also known as colour centres (*F* centres)[53] shows three dissimilar charge states, as $F^{2+}$ (unoccupied), $F^+$ (singly occupied), and $F^0$ (doubly occupied), in the ZnO lattice. In case of $F^{2+}$ and $F^0$ vacancies have spin-zero ground states; hence they do not induce FM in ZnO[54, 55]. However in case of $F^+$ vacancy which is singly occupied can contributes in activate bound magnetic polarons (BMPs) and magnetic moment in DMSs[54]. These defects arising due to Ni doping in ZnO nanostructure play a vital role in deciding the magnetic behaviour of the samples as has been discussed below.

It is very well recognised that for the extensive range of applications, the DMS materials should have a $T_C$, well above RT (300 K). The magnetization *vs* magnetic field (M-H) plots are shown in fig.8 (*a* and *b*), measured at 80K and for the ZnO:Ni 5% sample measured at 80K, 150K and 300K respectively (inset show room temperature M-H of pure ZnO).

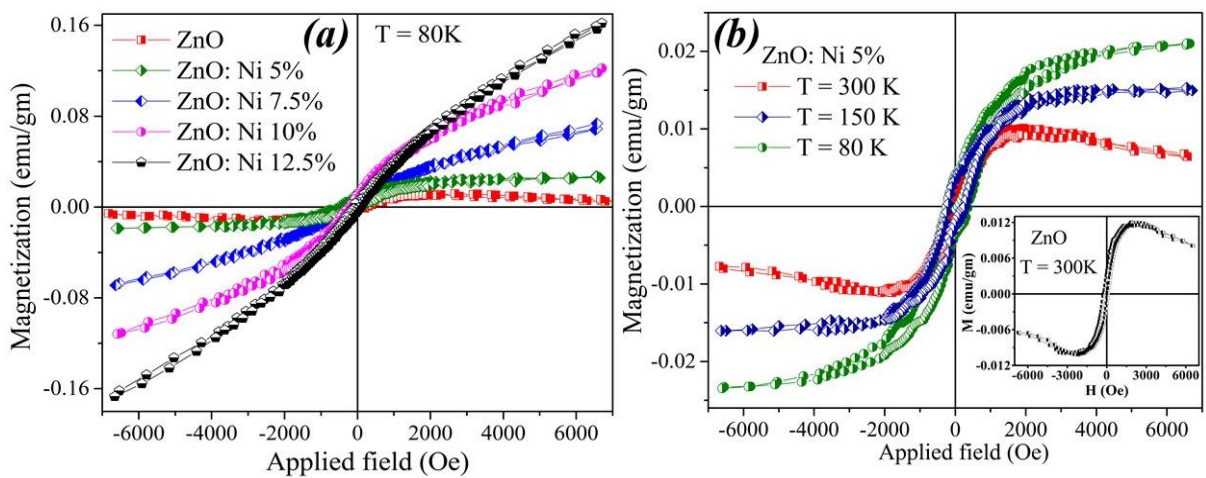

**Figure 8.** (**a**) *M–H plot at 80 K showing the hysteresis loop for pure and Ni doped ZnO nano structure and* (**b**) *Shows the M-H plot of Ni 5% doped ZnO at 80K, 150K and 300K and inset shows M-H plot of ZnO at 300K.*

Our experimental results of M-H curves (at 80 K) clearly shows the typical week ferromagnetic (FM) saturation behaviour of all the samples and saturation magnetization is found to increase from 0.02 to 0.17 *emu/gm.* with increase Ni concentration. It can be noted from *fig.* 8*(a)*, the saturation tendency is not increasing with increasing the Ni concentration however; in *fig.* 8*(b)* shows that, a decrease in the temperature does not alter the important FM characteristic of the samples but significantly change the saturated magnetic moments ($M_s$) of the samples.

It's still controversial debate about the origin of RT ferromagnetic property in Ni doped ZnO. There are four possible reasons generally proposed for the ferromagnetic behaviour of the samples *viz. (i)* Ni precipitation and the formation of NiO, *(ii)* at nanoscale level, formation of Ni-related secondary phase, *(iii)* extended defect due to the doping and *(iv)* presence of magnetic impurity (due to doped material) in the samples. The possibility of Ni precipitation or formation secondary phase NiO in the present set of samples can be easily ruled out, we did not find any evidence of NiO secondary phase in our sample from XRD or XANES result. Again the formation Ni related secondary phase at nanoscale level is also ruled out because other measurement such as EXAFS, PL, and Raman spectroscopy suggested magnificently incorporation of Ni ions at Zn sites of ZnO. Since all above measurements manifest integration of $Ni^{2+}$ ions in ZnO wurtzite and the formation of singly ionized oxygen vacancy, this can be considered to be the most important factor for FM nature of the present samples. Thus, TMs as well as defects produced by TMs, especially single ionized oxygen vacancy play the significant role to observed FM.

However for intrinsic ferromagnetism in TM doped ZnO, the precise mechanism is still not established beyond doubts. A number of different theories and ideas have been suggested for this, for example *(i)* the mean-field Zener model[5], *(ii)* RKKY *(Ruderman–Kittel–Kasuya–Yosida)* mechanism, *i.e.* carrier induced ferromagnetism *(iii)* donor impurity band exchange model, in this the FM in DMSs is due to an indirect exchange through shallow donor electrons and this form an BMPs[55-57] and *(iv)* direct interactions[6] (such as double or super exchange mechanism). RKKY is based on free electron, since ZnO is semiconductor in nature and small doping cannot transform it into metal, so the RKKY is invalid here. Again, direct interactions between the magnetic ions required for the double exchange mechanism are also not possible due to very dilute doping of the samples.

This suggests the intrinsic exchange interactions due to oxygen related defect which assisted BMPs and origin for the RT- FM in present case. Therefore we conclude that $Ni^{2+}$ ions cannot just produce intrinsic ferromagnetism unless and until there are formation of point defects such as singly ionized oxygen vacancies assisted BMPs for room temperature FM.

## 4. Summary and Conclusion

Ni-doped ZnO nanorods and nanoflakes have been successfully synthesized by simple wet chemical technique. Reitveld refined XRD patterns point out that for even up to 12.5% of Ni doping in ZnO the samples have no secondary phase related to Ni or NiO and shows the decrease in Zn/O occupancy ratio, which is associated with the defect induced due to Ni doping. Degree of distortion and lattice strain calculated from Williamson-Hall method, confirms the incorporation of Ni-ions without change in wurtzite structure of ZnO. SEM images show rod-like morphology in pure ZnO and with Ni doping the morphology of ZnO is changed from rods to flakes. The average length and diameter of nanorods in pure ZnO is ~2 $\mu m$ and 100 nm, respectively. XANES spectra confirm that Ni has $2^+$ oxidation state in Ni doped ZnO nanocrystals while it has been found from EXAFS analysis that there is an increase in disorder systematically and decrease in coordination number near the host (Zn) sites in the samples due to Ni doping. Raman study validates the results obtained from XRD studies suggests the doping of Ni ions without change in the crystal structure the ZnO lattice. Furthermore Raman spectroscopy shows that, with Ni doping there is shifting and broadening of LO ($A_1 + E_1$) peak with additional peak at 276 $cm^{-1}$, this confirms the presence of oxygen related defect such as $V_O$. This is further confirmed by PL spectra, showing peak at 524 nm associated to green emission, which is crosspsonding to singly ionized oxygen vacancies ($V_O^{+}$). These defects help to generate the BMPs in ZnO nanostructure. M-H plot shows the room temperature ferromagnetism, enhances with increase in the Ni concentration. From the above complementary studies it is thus proposed that joint effects of the Ni-ions as well as the intrinsic exchange interactions arising from defect such as $V_o^{+}$ assisted BMPs are accountable for the RT- FM in $Zn_{1-x}Ni_xO$ ($0<x<0.125$) system.

## Acknowledgments

This work is supported by the SERB-DST (Department of Science and Technology India) by awarding 'Ramanujan Fellowship' to the PMS (SR/S2/RJN-121/2012). PMS also acknowledge the CSIR research grant No. 03(1349)/16/EMR-II. We are thankful to Director Prof. Dr. Pradeep Mathur IIT Indore, for motivating the research and SIC-IIT Indore for providing basic characterization facilities. We are thankful to Vasant Sathe, UGC-DAE, Indore, and Dr. Rajasree Das, for their help to do Raman measurement and magnetic measurement, respectively.